\documentclass[useAMS,usenatbib,letters]{mn2e} 
\usepackage{graphicx} 
\usepackage{natbib}

\title[Cosmic web filaments and velocity field]{Orientation of cosmic web filaments with respect to the underlying velocity field}

\author[E. Tempel, N.~I. Libeskind, Y. Hoffman, L. J. Liivam\"{a}gi and A.~Tamm]
{E. Tempel$^{1,2}$, N.~I. Libeskind$^{3}$, Y. Hoffman$^{4}$, L. J. Liivam\"agi$^{1}$ and A.~Tamm$^{1}$\\ 
$^{1}$Tartu Observatory, Observatooriumi~1, 61602 T\~oravere, Estonia\\ 
$^{2}$National Institute of Chemical Physics and Biophysics, R\"avala pst 10, Tallinn 10143, Estonia\\
$^{3}$Leibniz-Institut f\"ur Astrophysik Potsdam, An der Sternwarte 16, D-14482 Potsdam, Germany\\ 
$^{4}$Racah Institute of Physics, The Hebrew University of Jerusalem, Givat Ram, Israel}

\begin{document}

\date{Accepted 2013 September 15.  Received 2013 August 31; in original form 2013 July 4}

\pagerange{\pageref{firstpage}--\pageref{lastpage}}
\pubyear{2013}

\maketitle

\label{firstpage}

\begin{abstract}
	The large-scale structure of the Universe is characterised by a web-like structure made of voids,
sheets, filaments, and knots. The structure of this so-called cosmic web is dictated by the local
velocity shear tensor. In particular, the local direction of a filament should be strongly
aligned with $\hat{e}_3$, the eigenvector associated with the smallest eigenvalue of the tensor. That conjecture is tested here on the basis of a cosmological simulation. The cosmic web delineated by the halo
distribution is probed by a marked point process with interactions (the Bisous model), detecting
filaments directly from the halo distribution (P-web). The detected P-web filaments are found to be strongly aligned
with the local $\hat{e}_3$: the alignment is within 30\degr\ for $\sim$80\% of the elements. This
indicates that large-scale filaments defined purely from the distribution of haloes carry more than
just morphological information, although the Bisous model does not make any prior assumption on the
underlying shear tensor. The P-web filaments are also compared to the structure revealed from the
velocity shear tensor itself (V-web). In the densest regions, the P- and V-web filaments overlap well (90\%),
whereas in lower density regions, the P-web filaments preferentially mark sheets in the V-web.
\end{abstract}


\begin{figure*}
	\includegraphics[width=58mm]{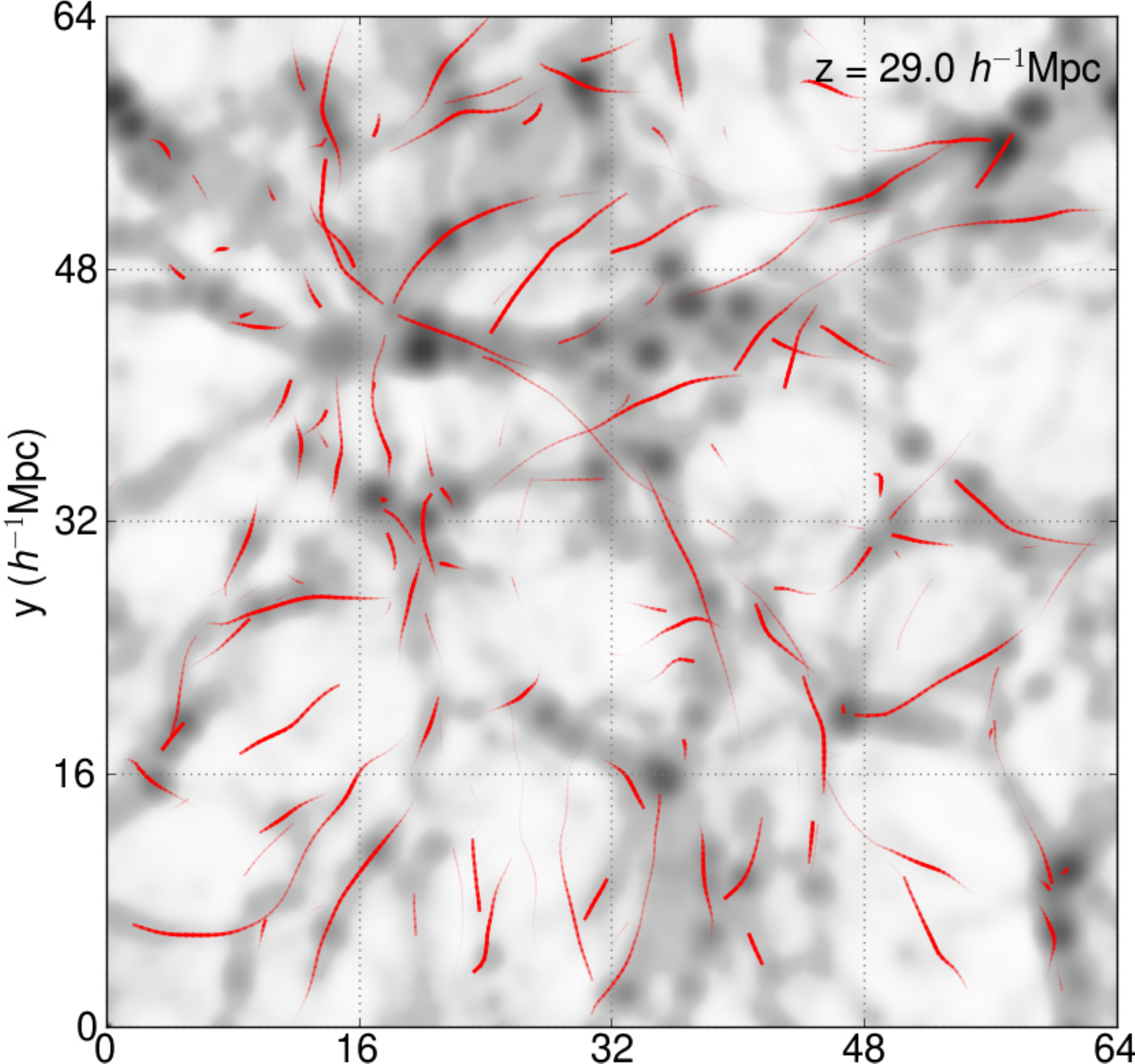}
	\includegraphics[width=58mm]{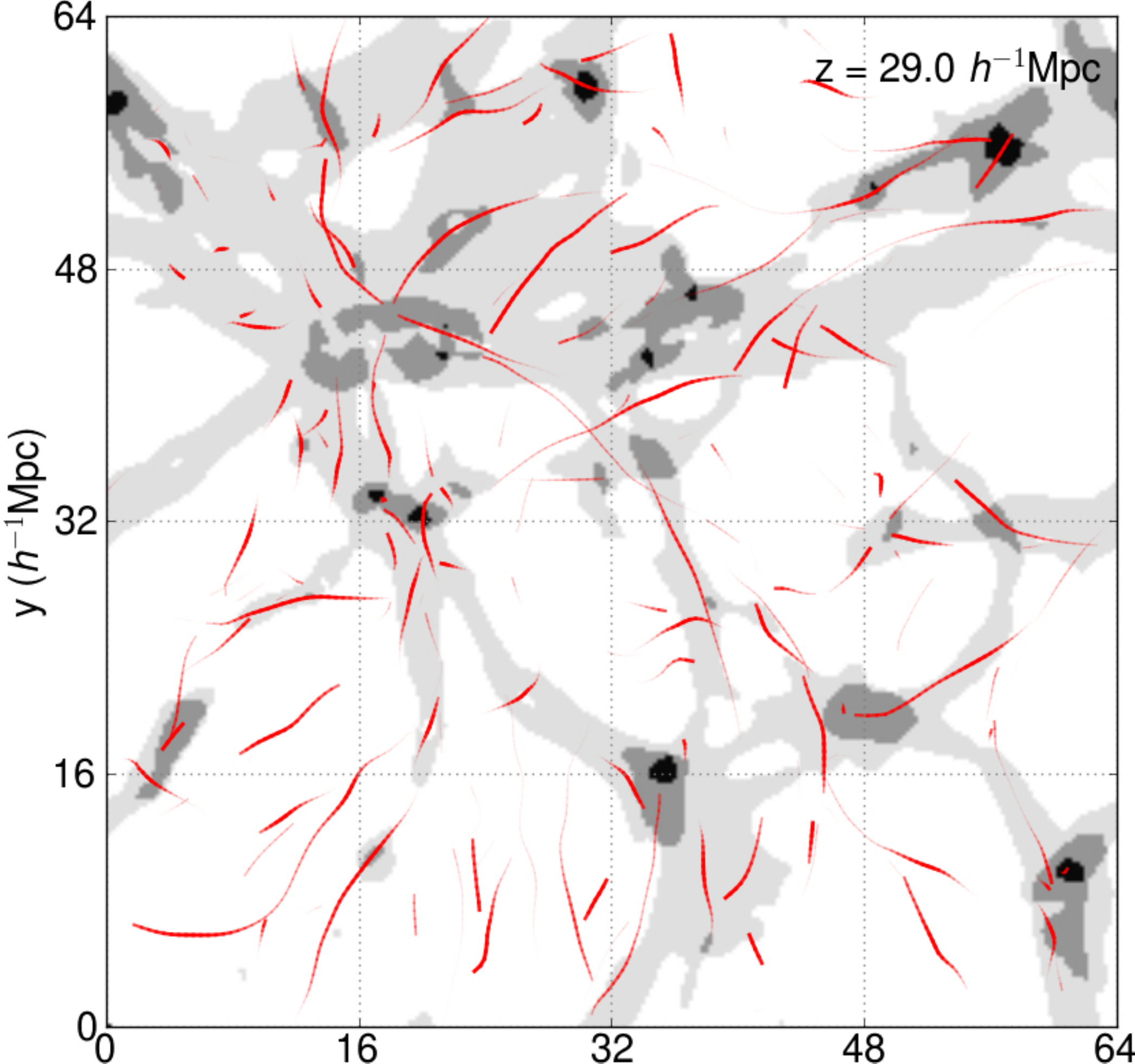}
	\includegraphics[width=58mm]{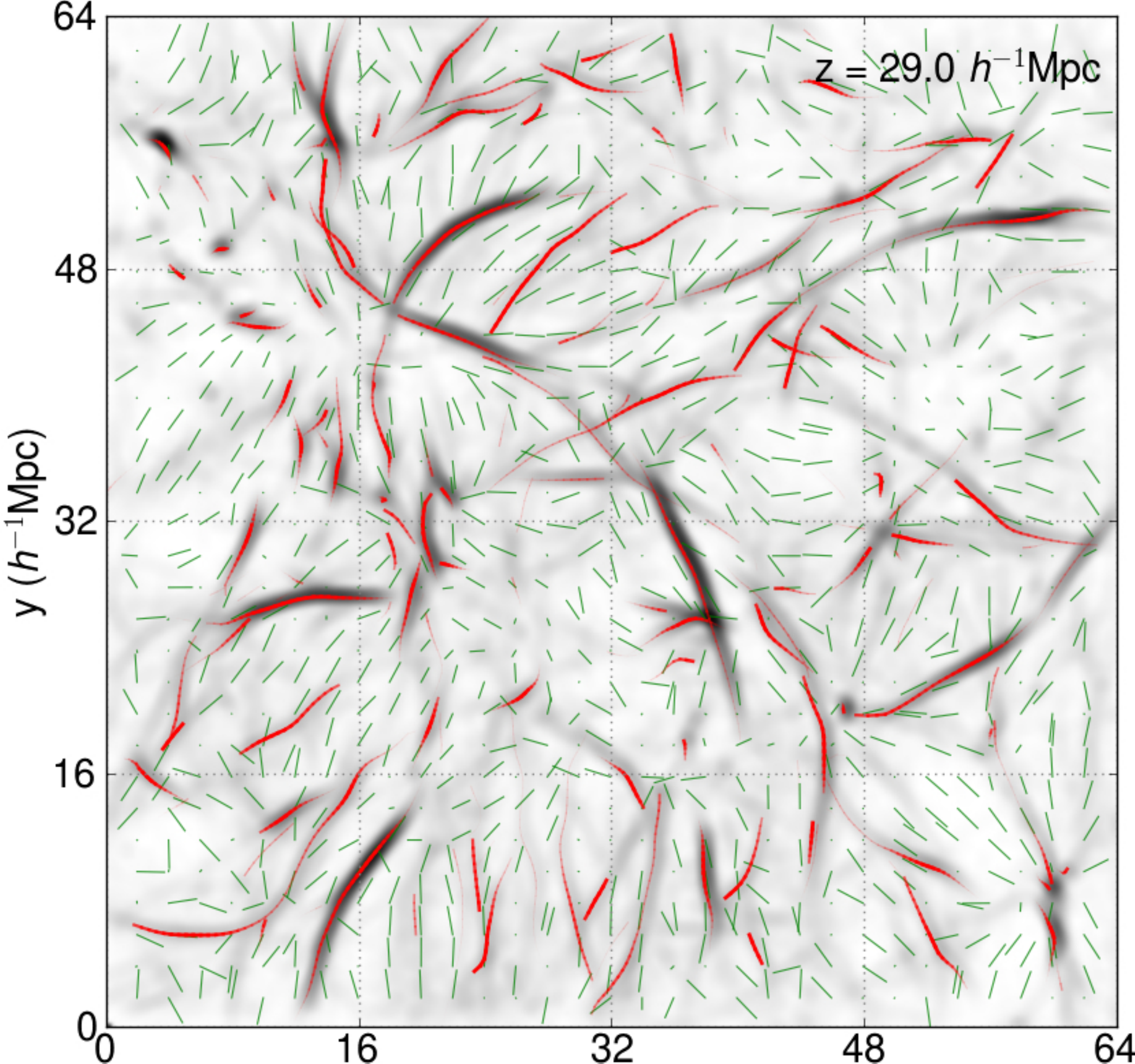} \\
	\includegraphics[width=58mm]{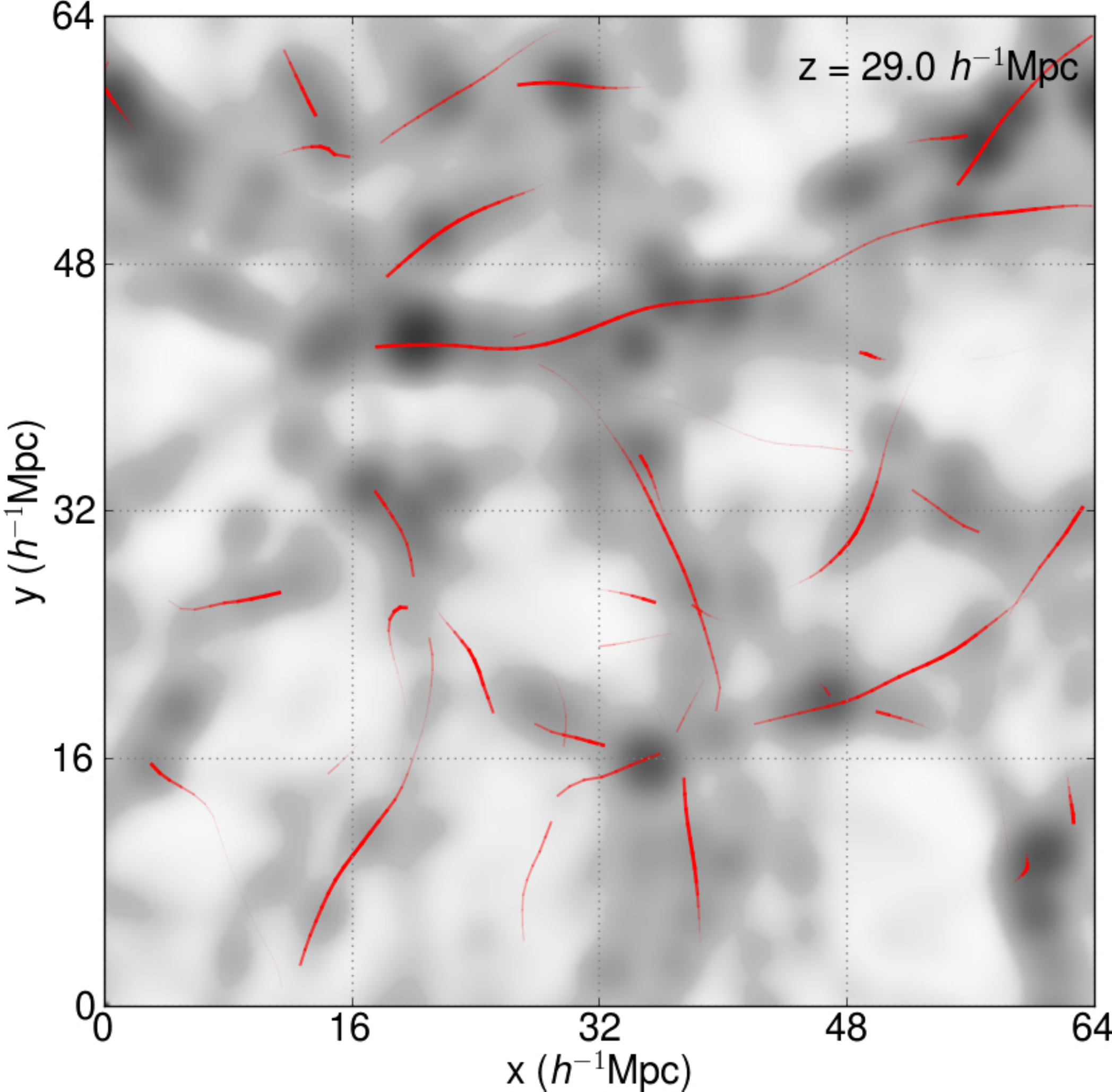}
	\includegraphics[width=58mm]{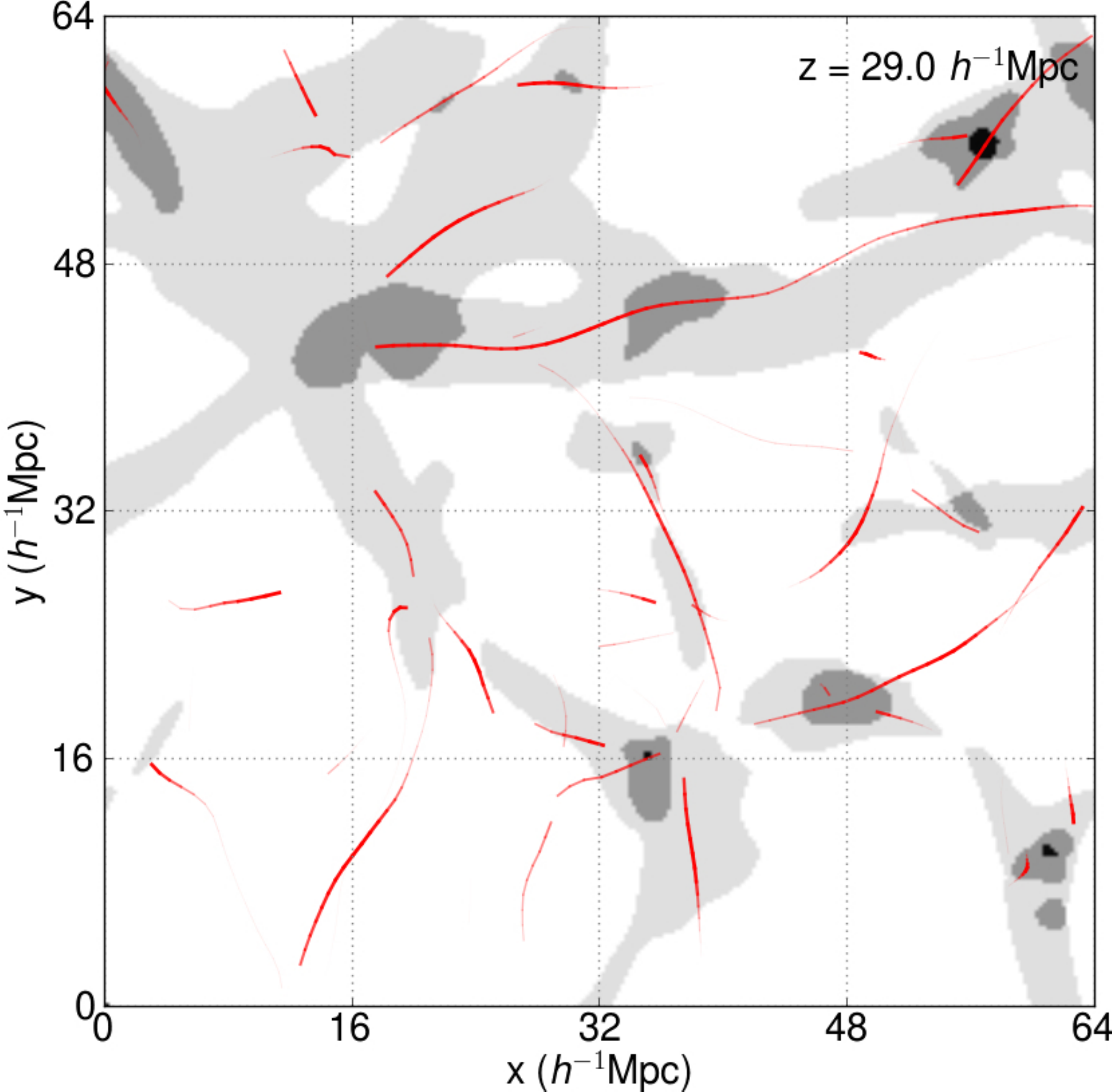}
	\includegraphics[width=58mm]{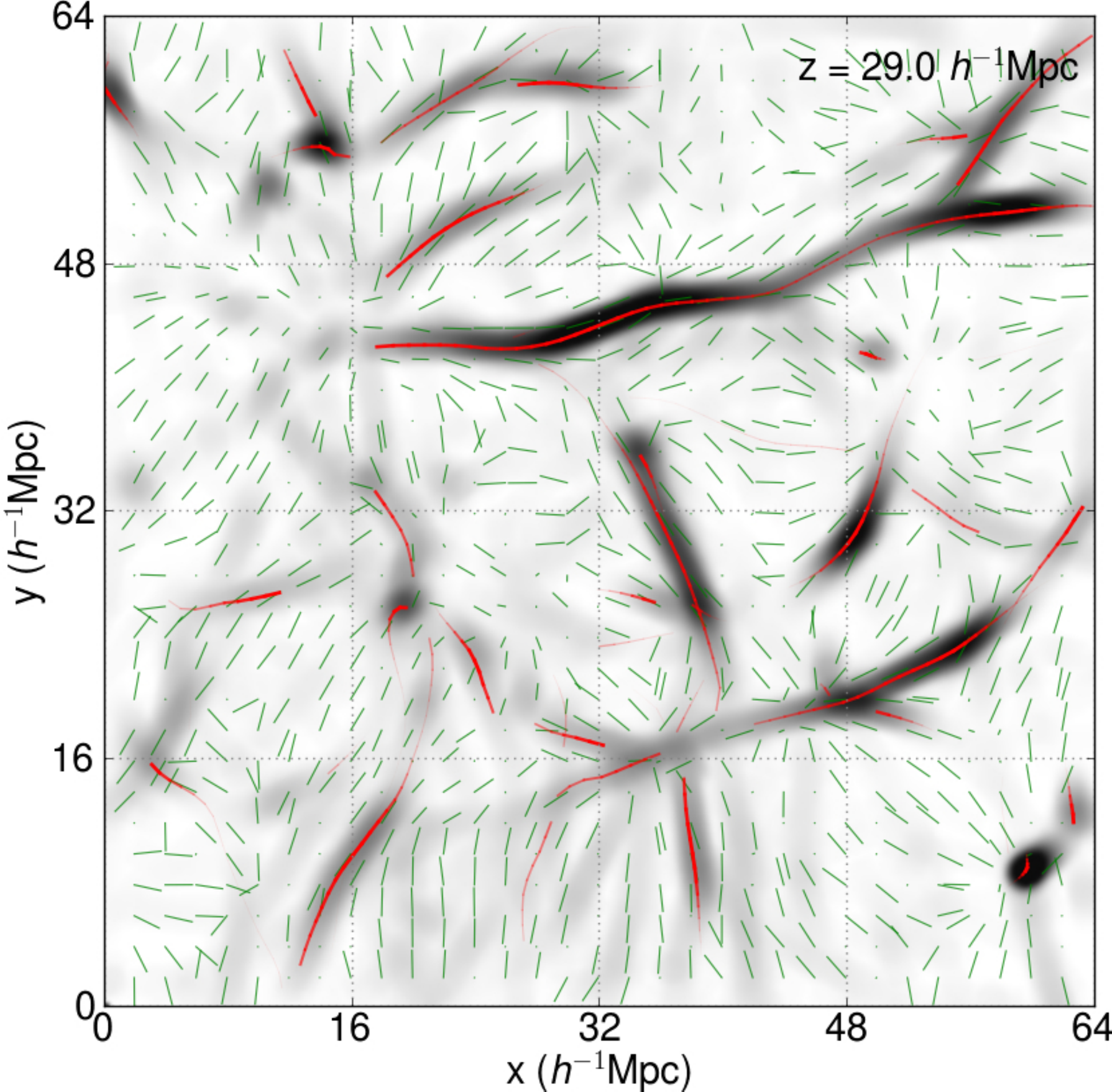}
\caption{The normalised density field overplotted with the P-web filament axes from point process 
(\emph{left panels}); web elements in the V-web (white for voids, light grey for sheets, dark grey for
filaments, and black for knots) based on velocity field overplotted with the same filament axes
(\emph{middle panels}); filament detection probability in Bisous process overplotted with projected
$\hat{e}_3$-vector of velocity shear tensor (\emph{right panels}). Upper row shows the filaments with
$r=0.5~h^{-1}\mathrm{Mpc}$ and lower row shows the filaments with $r=1.0~h^{-1}\mathrm{Mpc}$. Smoothing of the velocity and density field is roughly the same as the filament radius. The figure
represents one slice of the full computational box; the whole computational box can be seen in this
fly-through movie at \texttt{http://www.aai.ee/$\sim$elmo/PV-web/}.} 
\label{fig:maps}
\end{figure*}

\section{Introduction}

The large-scale matter distribution in the Universe represents a complex network of structure elements
such as voids, filaments, sheets and knots, forming the so-called cosmic web
\citep*{Joeveer:78,Bond:96}. The cosmic web has attracted the attention of both observers and
theoreticians and numerous studies have attempted to provide a quantitative description of the cosmic
web, trying to translate the visual impression into rigorous mathematical algorithms. Motivation for
constructing such methods ranges from the search for mathematical measures of the large scale structure
that can be used as a discriminator between alternative cosmological models to the desire for a
framework within which the environmental dependence of structure formation can be studied.

Currently, the studies concentrating on the large-scale environmental effects usually make implications
simply from the density field \citep[e.g.,][]{Blanton:05,Tempel:11,Lietzen:12}, while various
indications argue for a more intricate connection \citep{Lee:08}. For example, it is known that the spin
of dark matter (DM) haloes is correlated with the underlying web elements
\citep[e.g.][]{Codis:12,Libeskind:12,Libeskind:13} and there is observational evidence for the alignment
of the rotation axes of galaxies along galaxy filaments \citep[e.g.][]{Lee:07,Tempel:13}. A more
thorough insight into these relations requires definition algorithms for the large-scale structure.

Broadly speaking, web classifying algorithms follow one of two main streams. One is a classification
based on the point process manifested by the distribution of galaxies or clusters of galaxies, treated
as point objects. The other is based on the dynamics of the underlying density field and the velocity
field it induces. This was pioneered by \citet{Hahn:07} who used the tidal tensor of the underlying mass
distribution to classify the web \citep[see also][]{Forero-Romero:09}. \citet{Hoffman:12} and
\citet{Libeskind:12} classified the cosmic web by studying the velocity shear tensor of the underlying
mass distribution, by looking at the number of the eigenvalues of the shear tensor above a threshold.
This so-called V-web algorithm has been shown to improve the dynamical resolution with respect to the web
based on the tidal tensor, enabling the classification of structures on the scale of few tens of
kiloparsecs. The V-web has also been used by \citet{Libeskind:13b} to quantify vorticity (the anti-symmetric
part of the velocity deformation tensor) and its relationship to halo spins. A number of papers
have dealt with the myriad other ways (principally geometric, rather than dynamical) to characterise the
web \citep[e.g. see][]{Novikov:06,AragonCalov:07,Sousbie:11,Shandarin:13,Cautun:13}.

The Bisous model \citep[i.e. the marked point process with interactions,][]{Stoica:05} provides a
powerful tool for the construction of a network of filaments from the distribution of galaxies. The
algorithm operates directly on the point-like distribution of galaxies, or their DM halo counterparts,
without any explicit reference to the underlying dynamics in general, and the velocity shear tensor in
particular. (From here on it is dubbed as the P-web). The V-web's working hypothesis is that the shear
tensor is the main driver that shapes the cosmic web. In particular, the model firmly predicts that the
direction of filaments should be strongly aligned with the direction of $\hat{e}_3$, the eigenvector
corresponding to the smallest eigenvalue of the shear tensor. This is a local relation between the two
directions, namely it should be obeyed at any position on the filaments.

The aim of our paper is to test the filament-$\hat{e}_3$ hypothesis by studying a
high-resolution DM-only $\Lambda$CDM simulation. The simulation will be probed by a halo finder, and the
detected haloes will in turn be probed by a P-web finder. The velocity shear tensor is constructed on a
cartesian grid by Clouds-in-Cells interpolation scheme; the tensor is diagonalised on each grid
cell. The alignment of the filaments and the local direction of $\hat{e}_3$ are then compared.

A side project pursued here is to compare the filaments constructed by the P- and the V-web. However, a
comment of caution is due here. The cosmic web is an ill-defined structure. A close visual inspection
reveals a smooth transition from knots to filaments, from filaments to sheets, and eventually from
sheets to voids. There is no clear-cut principle that separates these classes and virtually
all web finders have one or more free parameters that dictate the transition from one class to
the other. Given the very different nature of the P- and the V-web algorithms one should not expect a
high level of overlap of the two kinds of filament, even if they share a common alignment.

\section{Data and methods} \label{sect:data}

\subsection{$N$-body simulation}

A DM-only $N$-body cosmological simulation is run assuming the standard $\Lambda$CDM concordance
cosmology \citep[e.g. WMAP5,][]{Komatsu:09}, in particular a flat universe with cosmological constant
density parameter $\Omega_\Lambda= 0.72$, matter density parameter $\Omega_\mathrm{m} = 0.28$, a Hubble
constant parameterised by $H_0 = 100~h~\mathrm{km}~\mathrm{s}^{-1}\mathrm{Mpc}^{-1}$ (with $h = 0.7$),
a spectral index of primordial density fluctuations given by $n_s = 0.96$, and mass fluctuations given
by $\sigma_8 = 0.817$.

The simulations span a box of side length $64~h^{-1}\mathrm{Mpc}$ with $1024^3$ particles, achieving a
mass resolution of $\sim\!1.89 \times 10^7~h^{-1}M_\odot$ and a spatial resolution of
$1~h^{-1}\mathrm{kpc}$. The publicly available halo finder AHF \citep{Gill:04,Knollmann:09} is run on
the particle distribution to obtain a halo catalogue. AHF identifies haloes and subhaloes in the
simulation by searching the particle distribution for local density by maxima and checking that
particles within the virial radius are gravitationally bound to the host structure. Substructures are
identified as haloes whose centres are located within the virial radius of a more massive parent halo.
Only haloes more massive than $10^9~h^{-1}M_\odot$ are considered in this work.

The initial conditions of the simulation were constrained using the Hoffman-Ribak algorithm
\citep{Hoffman:92} to reproduce the correct environment (on scales of $\sim$10\,Mpc) of the local
group \citep[see][]{Libeskind:10,Libeskind:11}.

\subsection{Velocity shear tensor -- the ``V-web''} 

As mentioned in the introduction, the cosmic web can be quantified by means of the velocity shear
tensor. This method is described in detail in \citet{Hoffman:12}, \citet{Libeskind:12,Libeskind:13}. The
salient aspects are highlighted here, in brief. The cosmic velocity field is calculated using a
``Clouds-in-Cell'' (CIC) algorithm on a $256^{3}$ grid. The cell size is thus 250~$h^{-1}$kpc and the
number of grid cells is chosen to be the finest mesh which ensures each cell contains at least one
particle. The velocity (and density) fields are then smoothed with a gaussian kernel equal to at least
one cell (i.e. $r_{\rm smooth}=250~h^{-1}$kpc) in order to get rid of the spurious artificial cartesian
grid introduced by the CIC. In practice the smoothing sets the scale of the calculation and we use two
smoothings throughout this paper: 500 and 1000~$h^{-1}$kpc. The velocity shear tensor is defined as
$\Sigma_{\alpha\beta}=\frac{1}{2H_{0}}\big(\frac{\partial v_{\alpha}}{\partial r_{\beta}}+\frac{\partial
v_{\beta}}{\partial r_{\alpha}}$\big) and is calculated by means of FFT in $k$-space. The velocity shear
tensor is then diagonalised and its eigenvectors and eigenvalues are identified. Note that the velocity shear field is identical to the tidal field, defined as the Hessian of the potential, namely $T_{\alpha\beta}=\frac{\partial^{2}\phi}{\partial r_{\alpha}\partial r_{\beta}}$, when smoothed on large enough (i.e. $>$ few Mpc) scales.

In addition to being a universal characteriser of the velocity field, the velocity shear tensor can be
used also for classifying the cosmic web. The eigenvector corresponding to the greatest eigenvalue of
the shear tensor denotes the direction along which material is collapsing fastest (or expanding
slowest), similarly for the intermediate and minor eigenvalues. In this way a web classification can be
carried out by simply counting the number of axes that are collapsing: 0, 1, 2 or 3 for voids, sheets,
filaments or knots, respectively. An axis is said to be ``collapsing'' if its eigenvalue is greater than
some threshold (chosen to be 0.5 so as to accurately reproduce the visual impression of the cosmic web).
Note that filaments are defined by two collapsing axes and an expanding one. The expanding axis thus has
the lowest eigenvalue and corresponds to the orientation of the filament, being identical to the
$\hat{e}_{3}$ vector of the V-field.

\begin{figure} 
	\centering
	\includegraphics[width=84mm]{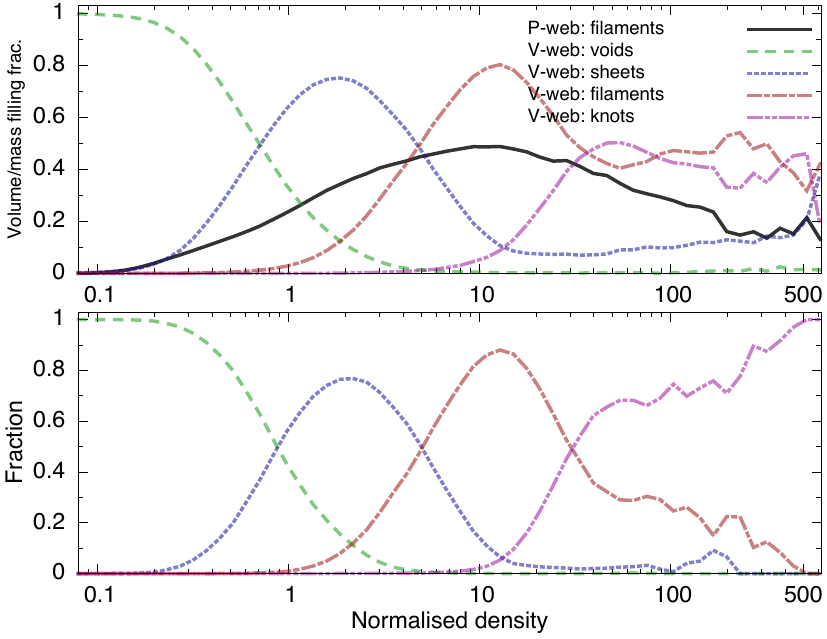} 
	\caption{\emph{Upper panel} shows the volume/mass filling fraction for V-web elements and P-web 
	filaments (black solid line) as a function of normalised density. P-web filaments are shown for
$r=0.5~h^{-1}\mathrm{Mpc}$ and the assumed volume coverage is taken to be twice of that. \emph{Lower
panel} shows the fraction of V-web elements for the P-web filaments as a function of normalised density.
Note that the fraction of V-web elements follows roughly the volume filling fraction of the same
elements. } 
\label{fig:frac_dens} 
\end{figure}

\subsection{Point Processes -- the ``P-web''}

We apply an object point process with interactions (the Bisous process) to trace the filamentary network
in the distribution of haloes. The method is applicable to an observed galaxy distribution as well
as dark haloes in simulations -- it just requires the spatial coordinates of the objects. The
morphological and quantitative characteristics of the initially complex geometry of the filamentary
network is obtained by sampling the probability density of the detected structures and by applying
the methods of statistical inference. We use the Metropolis-Hastings (MH) algorithm to sample the model
probabilities. A thorough explanation of the method can be found in \citet*{Stoica:07,Stoica:10}.
A realisation of the method used in this work is described in full detail in
\citet{Tempel:13b}.

In brief, the number density distribution of objects is probed by randomly placed segments (in the
present case by short thin cylinders). The likelihood of a cylinder is considered higher, if being
linked with another segment, thus forming an element of a filament.

After a large number of repetitions of the MH algorithm, a network of filaments emerges, where each
filament being labelled with coordinates, direction, and detection probability.
Altogether, we run 50 simulations and construct a filament detection probability field and the
orientation field. Based on these fields, we extract the filament axes as described in
\citet{Tempel:13b}. A movie, illustrating the Bisous process, the Metropolis-Hastings sampling, and the
detection of filament axes is available at \texttt{http://www.aai.ee/$\sim$elmo/PV-web/}.

The advantages of this method are manyfold. It is insensitive to observational systematics
introduced by survey geometry and selection effects and is capable of recovering poorly sampled
structures. Besides, since the Bisous model assigns a probability to each detected structure,
we are directly supplied with the reliability of the proposed filaments.

The method requires a fixed scale for the filament elements. In the present study, we seek filaments
at two scales: with radii in order of $r=0.5$ and $r=1.0~h^{-1}\mathrm{Mpc}$~--
the scales roughly corresponding to galaxy groups. The detected filaments can be seen as bridges
between galaxy clusters.

\section{Results} \label{sect:res}

\subsection{Filaments based on point process and velocity field}

In order to illustrate and quantify the P- and V-web filaments, the density field of a thin (2 and
4~$h^{-1}$Mpc for upper and lower panels, respectively) slice of the whole computational box is shown in
Fig.~\ref{fig:maps} (left panels), with a Gaussian smoothing with a 0.5 and $1.0~h^{-1}\mathrm{Mpc}$
kernel for the upper and the lower row, respectively. The detected P-web filaments are shown at two
scales: $r=0.5$ and $r=1.0~h^{-1}\mathrm{Mpc}$ filaments in the upper and the lower panels, respectively.
Visually, the filaments appear to delineate the underlying density field well. Some of the filaments are
detected at both scales, others not. In places where the density field exhibits broader structures, only
filaments of the scale $r=1.0~h^{-1}\mathrm{Mpc}$ are detected and vice versa.

The middle panels in Fig.~\ref{fig:maps} show the web elements as found by the V-web and the axes of
the filaments detected in the P-web. The size of the smoothing kernel of the velocity field
corresponds to the scale of the P-web filaments and the smoothing radius of the density field: 0.5
and $1.0~h^{-1}\mathrm{Mpc}$ for upper and lower panels, respectively. We note that in some cases,
the filaments detected from the V-web and the P-web are similar, but not always.
Comparing the V-web elements from the two different smoothings, the filamentary web looks
surprisingly similar, with differences emerging only at small scales. In contrast, differences between the P-web filaments are much larger. For
example, the most prominent filament in this slice of the simulation is visible at both V-web
scales, but is too fat to be detected as a $r=0.5~h^{-1}\mathrm{Mpc}$ P-web filament. Thus, while
V-web elements do not know the scale of the probed cosmic web element, the P-web elements correspond well to the actual scale
of a given structure.

\begin{figure} 
	\centering
	\includegraphics[width=84mm]{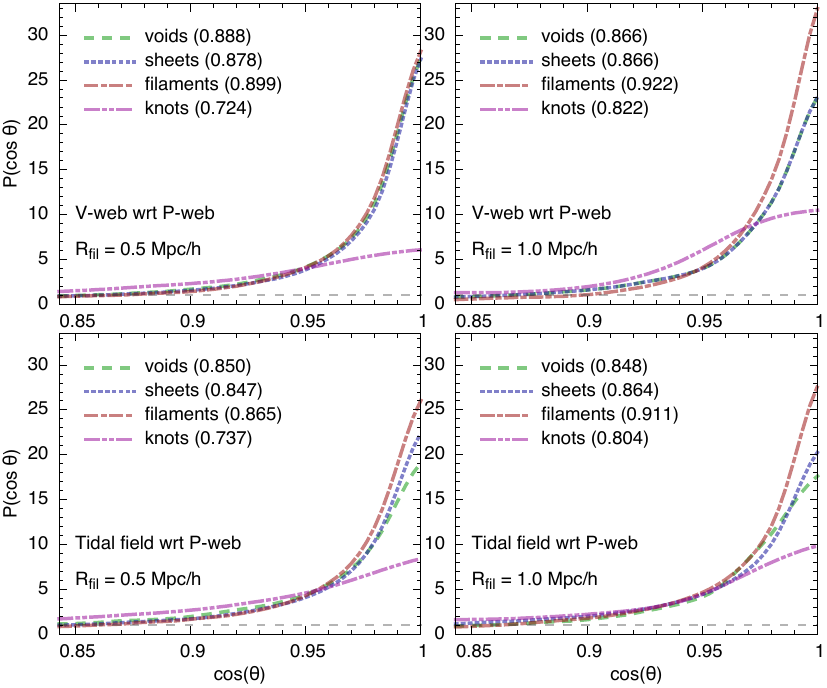}
	\caption{Distribution of $\cos{\theta}$ between the P-web filament axes and underlying 
	velocity field
(upper panels) and tidal field (lower panels). Distributions are shown for various V-web
elements. Grey dashed line shows the uniform distribution. \emph{Left panels} show the correlation
between $r=0.5~h^{-1}\mathrm{Mpc}$ P-web filaments and corresponding velocity/tidal
field; \emph{right panels} show the same for $r=1.0~h^{-1}\mathrm{Mpc}$ P-web filaments.
The average $\cos{\theta}$ is given in brackets.
} 
\label{fig:cor_web} 
\end{figure}

The upper panel in Fig.~\ref{fig:frac_dens} shows the volume/mass filling fraction for the V-web
elements and the P-web filaments. Since the plot is shown as a function of density, the volume and
mass filling fractions are one and the same. We see that from V-web, filaments are more likely
detected in higher density regions, as illustrated also by \citet{Libeskind:12} and
Fig.~\ref{fig:maps}, whereas the chances of finding filaments with the P-web are much less dependent
on the underlying density. Interestingly, the maximum volume filling fraction of both classification
methods occurs at the same density level, namely $10\rho_{\rm mean}$, although filaments exist at
all density levels.

The lower panel of Fig.~\ref{fig:frac_dens} shows the fractions of the P-web filaments detected as
different kinds of V-web elements. We note that these fractions follow rather closely the
volume/mass filling fraction of the V-web elements. This is not surprising, since filaments in the
halo distribution are also found in lower density environments, where the fraction of V-web
filaments is very low. P-web filaments are detected also in low density environments because the
classification/detection based on point process is built to be independent of the number density and
the procedure seeks defined structures. In general, the volume/mass filling fraction of P-web
filaments is less density dependent, indicating that filaments with some fixed scale exist
everywhere.

The filament finder based on the point process also detects filaments that are located in V-web
sheets: if a sheet has a filament-like overdensity, the P-web will detect it as a filament. On the
other hand, since the transition from sheets to filaments is smooth, it is hard to make difference
between filaments and sheets using the velocity field alone.

The differences described above make it difficult to compare these two web classification methods
directly. Clearly, within density range of 4--30$\rho_{\rm mean}$, the filaments detected by the two
methods are roughly the same, and the maximal overlap is $\sim$90\%. At lower and higher densities, the
comparison is not possible, since the detected structures are of different nature. However, the
P-web filaments can be directly compared to the underlying velocity field.

\subsection{Correlation between filaments and velocity shear tensor}

Both the P-web and the V-web return directions which are directly related to the large scale
structure. In the P-web's case this is the direction of each filament. In the V-web's case this is the
direction of the eigenvectors of the shear tensor. In this section we compare the direction of
filaments defined by the P-web with that of $\hat{e}_{3}$, the axis of least collapse.

The right panels in Fig.~\ref{fig:maps} show the filament detection probability field that is based
on point process. The filament detection field is overplotted with the projected
$\hat{e}_{3}$-vector of velocity shear tensor. From this Figure, we see that the detected filaments
and the underlying velocity field are very strongly correlated. The correlation is visible from
low-density environments to high-density environments.

To quantify the visible correlation between the P-web filaments and underlying velocity field, in
Fig.~\ref{fig:cor_web} (upper row) we show the distribution of cosine of the angle between the P-web filament
axis and the orientation of $\hat{e}_{3}$ vector of velocity shear tensor. 
For that we use the CIC shells that are less than a filament radius away from the P-web filament axes.
Distributions are shown for various web types defined by velocity shear tensor classification. The
velocity shear tensor is very strongly aligned parallel to the filaments detected by P-web: the
alignment is within 30\degr\ for $\sim$80\% of the detected elements (excluding knots). The correlation
is roughly the same for voids, sheets, and filaments, being slightly larger for the latter. The
correlation strength is independent of the defined filament scale. The same correlation shows that
filaments from point process classified as voids, sheets or filaments by velocity field are dynamically
the same structures: they just live in different density environments. The filaments defined by the
P-web are therefore intimately related to the underlying velocity field as characterised by the shear.

The lower row in Fig.~\ref{fig:cor_web} shows the distribution between P-web filaments and the tidal field (the ``linear'' velocity shear tensor). The correlation for the tidal field is slightly weaker than for the V-web, showing that P-web filaments are dynamically stronger than linear perturbations predict.

\section{Summary and Conclusions} \label{sect:summ}

A cosmological DM $N$-body simulation has been used to compare the filaments detected from the halo
distribution to the velocity field. The so-called V-web algorithm, here used to probe the underlying velocity field, was initially developed for the purpose of cosmic web
identification in numerical simulations, while the P-web generation algorithm, functioning as a marked point process, was developed for the identification of
filaments in large sky surveys. Although employing completely different techniques, philosophies and
motivations, these two methods show some remarkable similarities. Most importantly, \textit{the direction of
filaments defined using point-processes applied to the halo distribution matches very well the
eigenvector of the velocity shear tensor corresponding to slowest collapse}. This is our main
result.

On the other hand, the intimate correspondence between filaments found in the two methods (using the velocity
shear tensor and the point process based on halo distribution) are somewhat different: filaments defined
from the velocity shear tensor are mostly located in higher density environments, while filaments in
point process are also found in lower density environments and are in general less density-dependent. In principle the V-web could be ``tuned'' to reduce this density dependency (by lowering
the dynamical threshold above which axes are considered to be collapsing) however this may introduce
other problems (such as thicker and fatter filaments which fill a greater fraction of the volume).
However, in the densest regions, where filaments dominate the V-web, the two methods overlap in up to
90\% of the cases. A visual inspection confirms that point-process filaments found in low density
environment are real structures, not artefacts, which is also confirmed by their excellent alignment with the velocity field.

An exact filament-by-filament comparison between the two classification methods requires a more in
depth study, which should take into account the hierarchical nature of these structures. For
example, the V-web simultaneously identifies all web types (knots, filaments, sheets, and voids)
while the point process is designed only to identify filaments, thus a more sophisticated
comparison is complex. That said, it should perhaps be considered as a success (and
sanity check) that both of these methods, developed with completely different techniques, aims, and
philosophies -- find similar objects with similar orientations.

\section*{Acknowledgments}

We acknowledge the ESF grants MJD272, SF0060067s08; NIL acknowledges a grant from the \textit{Deutsche Forschungs Gemeinschaft}. YH has been partially supported by the Israel Science Foundation (1013/12). The simulations were performed and analysed at the Leibniz Rechenzentrum Munich (LRZ), at the Barcelona Supercomputing Center (BSC), and at the Computing Center of University of Tartu.


\label{lastpage}

\end{document}